\documentclass[preprint2]{aastex}

\shortauthors{Meier et al.}
\shorttitle{Warm Molecular Gas in Dwarf Starbursts}

\begin{document}
\input{psfig.sty}

\title{Warm Molecular Gas in Dwarf Starburst Galaxies: CO(3-2) Observations}
\author{David S. Meier, Jean L. Turner and
Lucian P. Crosthwaite} 
\affil{Department of Physics and Astronomy, University of 
California, Los Angeles, CA90095--1562
\\email:meierd,turner,lucian@astro.ucla.edu}
\and
\author{Sara C. Beck}
\affil{Department of Physics and Astronomy, Tel Aviv University,
Ramat Aviv, Israel
\\email:sara@wise.tau.ac.il}

\begin{abstract}
Eight dwarf starburst galaxies have been observed with the Caltech
Submillimeter Observatory (CSO) telescope in the CO J= 3 - 2
transition. The galaxies observed are He 2-10, NGC 5253, NGC 1569, NGC
3077, Haro 2, Haro 3, II Zw 40 and Mrk 86; all but the last two are
detected.  The central regions of He 2-10 and NGC 5253 were mapped and
a CO(2-1) spectrum of NGC 5253 was obtained.  The error weighted mean
CO(3-2)/CO(1-0) ratio of the detected galaxies is 0.60$\pm$0.06, which
is virtually identical to what is found for starbursts in the nuclei
of nearby spirals, and suggests that the molecular gas is optically
thick, warm (T$_{k}~>~$20 K), and moderately dense ($n_{H_{2}}\sim
10^{3-4}~cm^{-3}$).  The CO(3-2)/CO(1-0) ratio peaks at or close to
the starburst in all cases.  CO emission does not appear to be
optically thin in these dwarfs, despite the low metallicity and
intense radiation fields, which is probably because in order for CO to
exist in detectable amounts it must be self-shielding and hence
optically thick.  Physical properties of the molecular clouds in these
dwarf starbursts appear to be essentially the same as nearby spiral
nuclei, with the possible exception that CO is more confined to the
cloud cores.
\end{abstract}

\section{Introduction} 

Dwarf galaxies differ in many characteristics from large star-forming
spirals.  Typically they are much smaller ($\leq$3 kpc), have lower
rotation velocities ($\leq$30 km/s), generally warmer ISMs (T$_{d}~>
30$K), solid body rotation, and lack density waves, presenting less
complicated star forming environments than spiral galaxies
\citep[eg.][]{H71, GH84, TT86}.  Also, dwarf galaxies are
compositionally different, appearing to have relatively high HI and
dark matter fractions \citep{TM81}, relatively little dust,
and lower metallicities.  The molecular content in dwarf galaxies is
particularly difficult to study, since CO(1-0) is usually quite weak
due to the low metallicities, and therefore probably not a good tracer
of H$_{2}$ \citep{IDVD86, VH95, W95}.

The relatively weak CO(1-0) emission in dwarf galaxies leads us to
search for a stronger line with which to study molecular gas.  The
J=3-2 transition of CO may be more easily detected in dwarf starbursts
since it is relatively easily excited in warm gas (E/k = 33.2 K).  Due
to its higher characteristic temperature and critical density
($n_{cr}~\sim 2\times 10^{3}~cm^{-3}$), the CO(3-2) transition may be
more sensitive to warm, dense gas directly involved in the starburst.
It also has a higher optical depth than the CO(1-0), which can make
it easier to detect if CO(1-0) is optically thin.  

A sample of nearby dwarf starbursts have been observed in the CO(3-2)
transition with the Caltech Submillimeter Observatory (CSO). Among the
questions addressed are: do molecular clouds in dwarf galaxies have
different physical properties than their higher metallicity
counterparts?  Is the CO(1-0) line weak in dwarf galaxies because it
is optically thin?  Is the CO conversion factor in dwarf starbursts
different from large spirals because of these physical conditions?

\subsection{The Galaxy Sample}

A sample of eight nearby (\mbox{$\stackrel{<}{_{\sim}}$}20 Mpc) dwarf
starburst galaxies was observed.  The galaxies are He 2-10, NGC 5253,
NGC 1569, NGC 3077, Haro 2, Haro 3, II Zw 40 and Mrk 86.  The sample
is heterogeneous and illustrative, but not complete or unbiased.
Galaxies were selected based on the following criteria: they have been
previously observed and detected in either the CO(1-0) or CO(2-1)
transition \citep*{WH89,BSH89,SSLH92,BIK94,GBJM96}, have IRAS
100$\mu$m fluxes greater than 5 Jy \citep{TT86, MI94}, are fainter
than M$_{B}~ > -18.5$ and have undergone a recent burst of intense
star-formation.  All the galaxies except NGC 3077 and Mrk 86 have
Wolf-Rayet (WR) emission features \citep*[eg.][]{C91, SSMRM96, GLHC97,
SCP99}.  Some of these galaxies contain super star-clusters (SSCs) (He
2-10, NGC 5253, NGC 1569 and Haro 3)\citep{CV94, MHLKRG95, AS85,
SSMRM96}.  A number of these dwarfs show signs of interaction (He
2-10, NGC 5253, NGC 1569, NGC 3077 and II Zw 40).  The properties of
the galaxies in the sample are shown in Table \ref{tbl1}.

\section{Observations}

We observed the $^{12}$CO(3-2) line (345.796 GHz) towards the galaxies
He 2-10, NGC 5253 and II Zw 40, on 1997 February 24 \& February 25,
and the remaining five galaxies on 1999 January 14 \& January 15 using
the 10.4m Caltech Submillimeter Observatory (CSO).  The beamsize of
the CSO at 345 GHz is 22$^{''}$.  Facility SIS receivers were used
together with a 1024 channel 500 MHz AOS spectrometer. System
temperatures ranged from 600 K - 1100 K for the 1997 data and 700 K -
900 K in 1999 ($\tau_{225} \simeq$ 0.08 - 0.15).  Pointing was checked
using IRC+10216, $\alpha$ Ori, (1997 observations) Saturn, and Mars
(1999).  Absolute pointing uncertainty was $\le$6$^{''}$, and
repeatable to $\leq 3-4 ^{''}$.  Reported temperatures are main beam
temperatures, which is the brightness temperature a source would have
if it uniformly filled the main beam and was zero elsewhere.  The main
beam efficiencies used to convert the antenna temperature to the main
beam temperature were determined from observations of Mars and found
to be $\eta_{mb}$ = 0.6 (1997) and 0.62 (1999).  Since no published
CO(2-1) spectrum exists for NGC 5253, G. Serabyn kindly obtained one
for us on 1997 May 28.  At 230 GHz, the CSO beamsize is 30$^{''}$, and
$\eta_{mb}$ = 0.65 as determined from observations of Mars.  System
temperatures for this spectrum are $\sim$400 K.  Second order or lower
polynomials were removed from the each scan to give a flat baselines.
The scans were then averaged, using 1/T$_{sys}^{2}$ weights.  The
spectra were Hanning smoothed to a resolution of 8.0 km s$^{-1}$
(1997) and 6.6 km s$^{-1}$ (1999).  Reduction of the spectra were done
in CLASS.  For He 2-10, the maps were made and analyzed using the NRAO
AIPS package.

\subsection{The CO(3-2) Spectra}

CO(3-2) spectra of the central positions of each of the galaxies are
displayed in Figure 1.  We detected CO(3-2) in all the galaxies except
Mrk 86 and II Zw 40.  Line intensities, uncertainties, and gaussian
fits are tabulated in Table \ref{tbl2}.  For some of the spectra
gaussians may not represent the true line profile but given the
weakness of the signal, fitting anything more complicated is
unwarranted.  The details of each galaxy are discussed in \S6.

\begin{figure}
\centerline{\psfig{figure=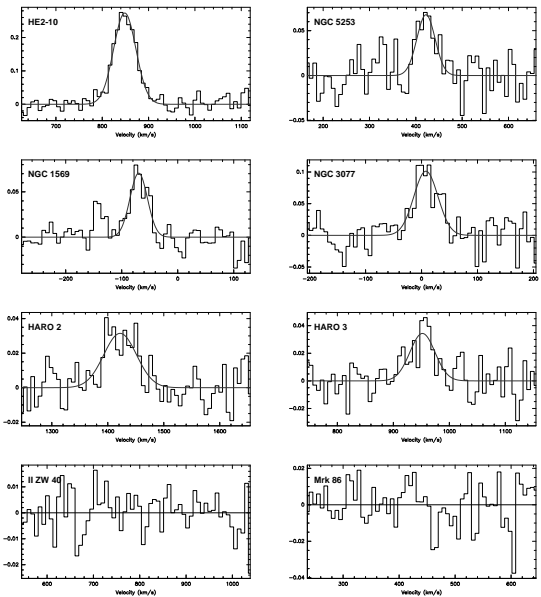,width=7.5cm,angle=00}}
\caption{The CO(3-2) spectra of the eight
galaxies. The vertical axis is main-beam temperature in units of K.
The horizontal axis is velocity (LSR) in units of km s$^{-1}$.  The
line represents the least-squares fitted gaussian.}
\end{figure}

In general CO(3-2) is very weak (T$_{MB}~\sim$ 0.05 K) over our
projected beamsizes of 0.25 - 2 kpc.  This is not surprising; dwarf
galaxies are known to be weak in the other low J CO lines
\citep*{YGH84, I86, TY87, ACCK88, SSLH92}.  However, CO(3-2) is strong
relative to CO(1-0) and CO(2-1).  Only one of the six detected
galaxies has CO(3-2)/CO(1-0) ratio as low as 0.4, which is typical of
the Galactic disk.  Comparing the CO(3-2)/CO(1-0) line ratio with the
CO(2-1)/CO(1-0) line ratio, we find that galaxies with high 2-1/1-0
ratios tend to have high 3-2/1-0 ratios.  There is one notable
exception, II Zw 40.  II Zw 40 has a high CO(2-1)/CO(1-0) ratio
\citep{SSLH92}, but its CO(3-2)/CO(1-0) ratio is $\leq$0.43
(1$\sigma$).

\section{Line Ratios: Physical Properties of the Molecular Gas}

CO line ratios can be used to investigate molecular gas properties in
these starbursts.  CO(3-2)/CO(1-0) ratios are more sensitive to gas
temperature than CO(2-1)/CO(1-0) ratios because of the larger
separation in J.  Thanks to the fortuitous matching of the CO(1-0)
beam at the IRAM 30m ($\sim$21$^{''}$) and the CO(3-2) beam at the CSO
($\sim$22$^{''}$), CO(3-2)/CO(1-0) line ratios can be estimated
without assumptions about source structure.  Only the two southern
galaxies, He 2-10 and NGC 5253, do not have IRAM CO(1-0) spectra.  For
these two galaxies, the source has been mapped at CO(3-2) or with an
interferometer at CO(2-1) or both \citep{MT98}.

Along with the CO(3-2)/CO(1-0) line ratios, Table \ref{tbl3} includes
CO(2-1)/CO(1-0) line ratios.  The CO(2-1) observations do not have
matching beams. Therefore Table \ref{tbl3} indicates what ratio would be
predicted in the point-source limit (source much smaller than either
beam) and in the uniform filling limit (source much larger than both
beams).  These line ratios are calculated using \citep[eg.][page 362]
{RW96}:
$$
R_{ij}~=~\frac{I_{21}}{I_{10}}\left(\frac{\Omega_{s} +~ 
^{21}\Omega_{B}}{\Omega_{s} + ~^{10}\Omega_{B}} \right)
$$
where $\Omega_{s}$, $^{21}\Omega_{B}$ and $^{10}\Omega_{B}$ are the
source solid angle and the solid angle of the beams at the respective
transitions, and $I_{21}$ and $I_{10}$ are their observed
line intensities.  For sources which have been mapped a deconvolved
ratio based on the measured $\Omega_{s}$ is also included.  In all
cases, with the possible exception of the CO(2-1)/CO(1-0) in NGC 1569,
the uniform filling ratios are unrealistically high, implying the
sources are small.  This is consistent with the low
filling factor found for these galaxies (\S3.1)

The CO(3-2)/CO(1-0) ratios in the dwarf starbursts range from 0.37 -
1.1.  The error weighted mean CO(3-2)/CO(1-0) line ratio for the
sample is 0.6 $\pm$0.06.  (We have estimated the uncertainty in the
line ratio as the noise error in each spectrum added in quadrature
with 20\% absolute calibration uncertainties.)  This error weighted
ratio is identical, within the uncertainties, to what is found for a
sample of non-dwarf starburst nuclei \citep[0.64;][]{DYSNY94} and
for nearby luminous IR galaxies \citep[$\sim$0.7;][]{MHWS99}.
The value is higher than the value obtained for Galactic GMCs, 0.4,
and closer to the value obtained for star-forming cores in the Galaxy
\citep[0.6;][]{STSWZ93}.

\subsection{LTE Modeling: Optically Thick CO Emission}

Comparing CO(3-2) to the other CO transitions (CO(3-2)/CO(1-0) line
ratio is used unless otherwise stated) can, in principle, be used to
constrain gas excitation temperatures and optical depths.  The ratio
of the CO(3-2) intensity to the CO(1-0) intensity, under the LTE
assumption is:
$$
\frac{I_{32}}{I_{10}}~=~{\int T_{32}dv \over \int T_{10}dv}~
$$
$$
\simeq~{f_{32}(J_{32}(T_{ex}) - J_{32}(T_{cmb}))(1-e^{-\tau_{32}})
\over f_{10}(J_{10}(T_{ex}) - J_{10}(T_{cmb}))(1-e^{-\tau_{10}})}
$$
where $J_{\nu}(T_{ex})~=~(h\nu/k)/(exp\{h\nu/kT_{ex}\}-1)$, with
$\tau$ and $f$ being the optical depth and filling factor of each
transition, respectively \citep[eg.][]{HTHM93}.  To increase the
signal-to-noise ratio (SNR), we have used the ratio of integrated
intensities instead of the peak main-beam temperatures.  This requires
that the CO(3-2) and CO(1-0) line profiles be similar.  For most of
the galaxies this is a good approximation (\S6).  For the LTE
approximation, ratios greater than unity indicate warm, optically thin
gas, while ratios less than unity indicate optically thick molecular
gas.

In practice, non-LTE effects such as different source sizes for the
CO(3-2) and CO(1-0) transitions or different T$_{ex}$ for each
transition due to temperature gradients and differential optical depth
effects in externally heated clouds may affect the ratio.  This
dataset cannot address such details. A discussion of these details
would require higher resolution or observations of rare CO
isotopomers.  The $^{13}$CO isotopomers, whose transistions will be
less optically thick, can be helpful in better constraining the
properties of these galaxies, but unfortunately these galaxies are too
weak in $^{13}$CO to detect with current telescopes (except He 2-10).

Previous studies indicate that variations in T$_{ex}$ become important
in localized regions of high column density, where optical depths of
order unity are reached over very small physical distances
\citep*[eg.][]{THH93, MTH00}.  When averaged over large single-dish
beams, the potential error associated with such effects appear to be
minor.  (For example, in the case of the nucleus of the nearby
metal-rich starburst, IC 342, a comparison of the physical conditions
derived using single-dish $^{12}$CO observations \citep{E90} are
similar to those obtained with higher resolution, even with the
presence of strong local temperature gradients \citep{MTH00}.  Similar
results have been found from studies of Galactic star-forming regions
\citep*[eg.][]{WHB99}.

Nearly all of the galaxies we observe have ratios \mbox{$\stackrel
{<}{_{\sim}}$}1.0 which indicate optically thick emission.  NGC 3077
(1.1) and the one velocity component of Haro 3 (1.3) have ratios that
are larger than one, but only marginally so.  In the case of NGC 1569
and NGC 3077, some portion of the high line ratios are due to a
slightly larger linewidth of the CO(3-2) line relative to the CO(1-0)
lines, indicating that there maybe some regions off line center that
have CO(3-2)/CO(1-0) ratios greater than 1.0.  But in both cases the
SNR is too low to say with certainty.  It appears that optically
thick gas dominates the CO emission in these dwarf galaxies.

We estimate the gas excitation temperatures of the six detected
galaxies, assuming the molecular gas is optically thick.  Derived
T$_{ex}$ range from $\sim$ 6 K (Haro 2) to $>$50 K (NGC 1569, NGC
3077, Haro 3).  For the high ratio galaxies a specific excitation
temperature cannot be obtained because the CO(3-2)/CO(1-0) ratio
ceases to differentiate temperatures well for $T_{K}$\mbox{$\stackrel
{<}{_{\sim}}$}30 K.  As a result of beam dilution, T$_{MB}$/T$_{ex}$
gives an estimate of the areal filling factor for each of the
galaxies.  Rather low filling factors are found, with values ranging
from $f_{a}$ $\sim$0.02 to $<$0.001 (Table \ref{tbl4}).

\subsection{LVG Modeling: Molecular Cloud Densities}

\begin{figure}
\centerline{\psfig{figure=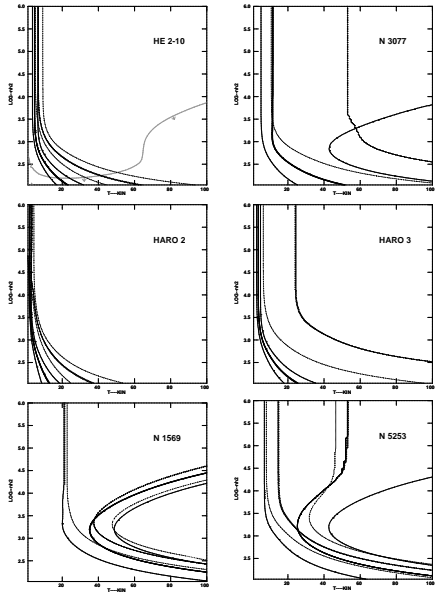,width=7.5cm,angle=00}}
\caption{The LVG models for the six detected
galaxies.  The solid line is for the CO(2-1)/CO(1-0) line ratio and
the dashed line is for the CO(3-2)/CO(1-0) line ratio.  For each case,
the bold line represents the measured value and the thin lines
represent the $\pm 1\sigma$ range.  The best fit solutions are the
regions where both line ratios overlap.  All models displayed here use
a velocity gradient of 1 km s$^{-1}$pc$^{-1}$.  The [CO/$H_{2}$]
abundances are $8\times 10^{-5}$ (1 Z$_{\odot}$) for He 2-10 and NGC
3077, $2.7\times 10^{-5}$ (1/3 Z$_{\odot}$) for Haro 2 and Haro 3, and
$1.1\times 10^{-5}$ (1/7 Z$_{\odot}$) for NGC 1569 and NGC 5253.  For
He 2-10, the $^{12}$CO(1-0)/$^{13}$CO(1-0) line ratio was also
modeled.  The grey curve represents the lower limit of Baas et
al. 1994 over the same beamsize as the CO(3-2) line ratios (and
consistent with the value obtained by Kobulnicky et al. (1995) over a
larger beamsize).  The abundance per unit velocity gradient used in
this model is $5\times 10^{-7}$ km s$^{-1}$pc$^{-1}$.}
\end{figure}

\begin{figure}
\centerline{\psfig{figure=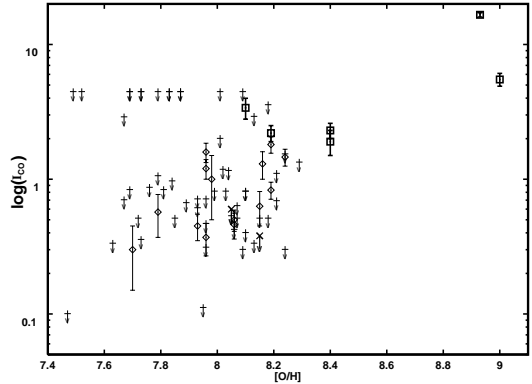,width=7.5cm,angle=-90}}
\caption{The CO(3-2) intensity in K km s$^{-1}$ is
plotted versus metallicity.  Bold open squares are the CO(3-2)
detections, and bold crosses are upper limits.  For reference, the
data is plotted over the dwarf galaxy CO(1-0) spectra compiled from
the literature by of Taylor et al. (1998).  Diamonds represent
detections and the arrows are upper limits. }
\end{figure}

\begin{figure}
\centerline{\psfig{figure=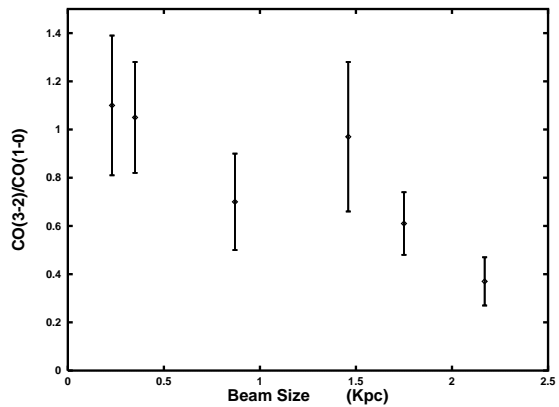,width=7.5cm,angle=-90}}
\caption{The CO(3-2)/CO(1-0) line ratio, as a
function of the projected physical size of the beam over which the
ratio has been measured, for the galaxies detected from our sample.  A
weak correlation is seen. }
\end{figure}

\begin{figure}
\centerline{\psfig{figure=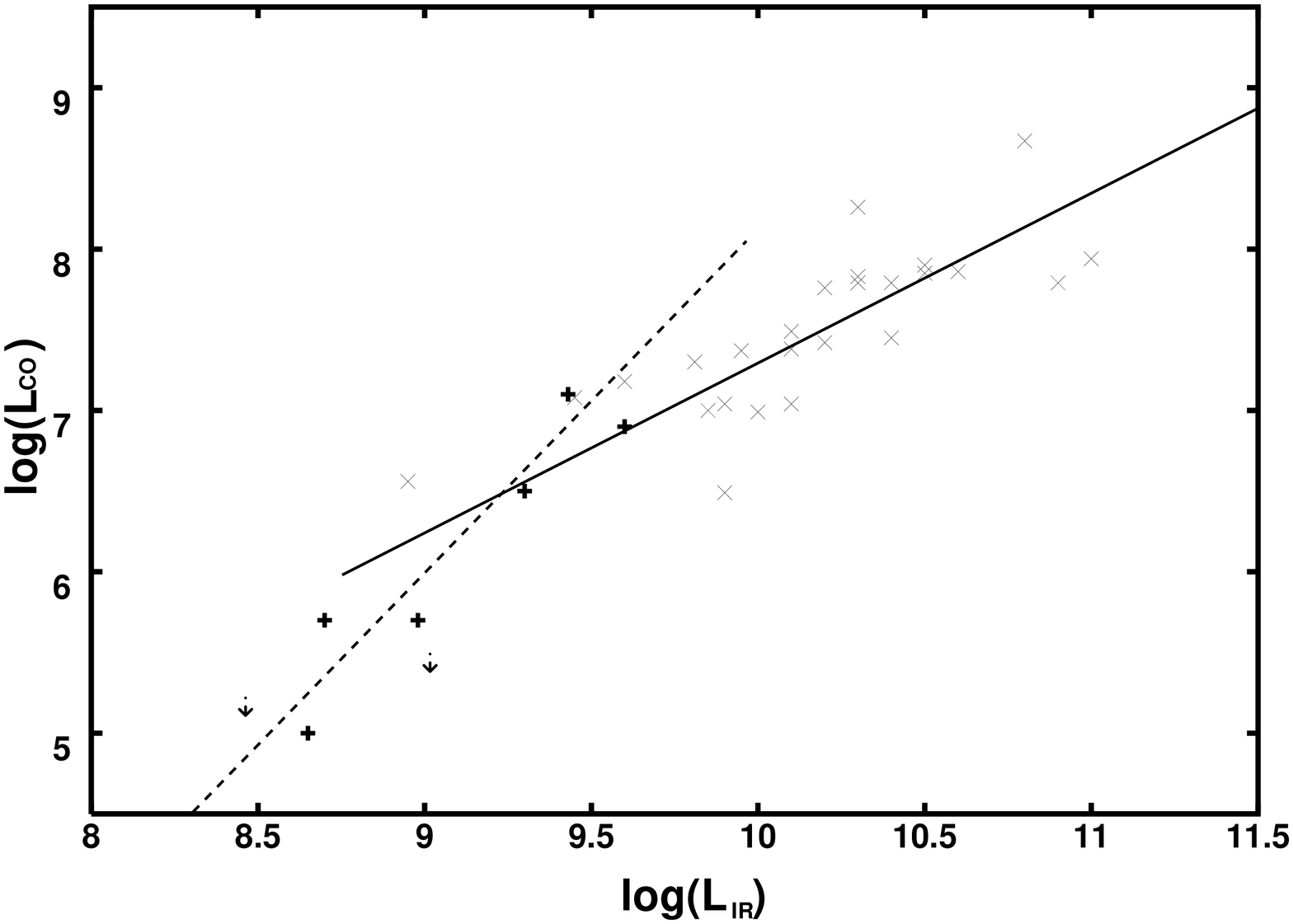,width=7.5cm,angle=-90}}
\caption{The luminosity of CO(3-2) in units of K km
s$^{-1}$pc$^{-2}$ plotted versus the infrared luminosity in units of
L$_{\odot}$, for the dwarf galaxies (bold pluses) and for a sample of
non-dwarf galactic nuclei (Mauersberger et al. 1998; crosses).  The
dashed line is the least-squares fit to the dwarf galaxy sample and
the solid line is the least-squares fit to the spiral nuclei sample.}
\end{figure}

Additional constraints on the physical conditions of the molecular gas
can be obtained using a Large Velocity Gradient (LVG) model
\citep*{GK74, DCD75}.  Due to the relatively large Doppler shifts
within the cloud, the emergent intensity can be related to its local
sources of excitation, T$_{k}$ and $N_{H_{2}}$.  As a result, LVG
models can be used to get an indication of the gas kinetic temperature
and density, given observations of several line ratios or antenna
temperatures.  While filling factors do not strongly effect the line
ratios for our matched beam observations, they do sensitively effect
the brightness temperatures.  The molecular clouds in these galaxies
are unresolved.  Therefore, only line ratios provide useful
constraints.  Detailed LVG modeling is not possible with the two line
ratios, CO(3-2)/CO(1-0) and CO(2-1)/CO(1-0), since there are three
unknowns, the gas density, $n_{H_{2}}$, the kinetic temperature,
T$_{K}$ and $X_{CO}$/$dv/dr$.  As mentioned earlier, temperature
gradients can also contribute; however, for a reasonable assumption of
the abundance per velocity gradient it is possible to provide rough
constraints on the physical parameters.  Given our limited knowledge
of the physical conditions in dwarf starbursts, even rough constraints
are important.

The LVG models were run for gas densities ranging from $n_{H_{2}} =
10^{2} - 10^{6}~cm^{-3}$ and kinetic temperatures T$_{K}~ 1 - 100$ K,
for a CO abundance per velocity gradient, $X_{CO}$/$dv/dr$, ranging
between $10^{-6}$ and $10^{-3}$.  Given that we only have two lines to
constrain three variables, model solutions are displayed for only one
choice of $X_{CO}$/$dv/dr$.  Figure 2 displays the best fit solutions
for T$_{K}$ and $N_{H_{2}}$ with an $X_{CO}$/$dv/dr$ estimated in the
following manner.  The CO abundance is assumed to be the solar CO
abundance scaled relative to the galaxies' metallicity and a constant
velocity gradient of 1 km s$^{-1}$pc$^{-1}$, which should roughly
account for metallicity effects \citep{VH95, W95}.  The solutions
obtained are relatively insensitive to $X_{CO}$/$dv/dr$ as long as it
is within a factor of 4-5 times the displayed value.  For different
assumed values of $X_{CO}$/$dv/dr$, the best fit solutions shift
slightly towards higher (lower) densities, for lower (higher) values
of $X_{CO}$/$dv/dr$, due to the rise of the effective critical density
caused by decreased radiative trapping.

\citet{SSLH92} argue that CO(2-1)/CO(1-0) line ratios obtained by
approximating the source size as a point source are closer to reality
because of the small CO source sizes expected for dwarf galaxies.  The
low derived CO(3-2) filling factors appear to confirm this. Therefore,
the ``point-source'' ratio approximations are used for galaxies with
no mapping information (Table \ref{tbl3}).  The details of the LVG
solutions for each galaxy are addressed in \S6, but the basic results
can be understood in simple terms.  In order to excite a bright
CO(3-2) line the gas must either be warm or have a relatively high
density.  At high densities the lower J transitions of CO will be
thermalized and the CO(2-1)/CO(1-0) and CO(3-2)/CO(1-0) values will
just reflect the Rayleigh-Jeans temperature corrections.  At low
densities, CO(3-2) weakens faster than CO(2-1) because of its higher
critical density.  Therefore, galaxies with similar CO(2-1)/CO(1-0)
and CO(3-2)/CO(1-0) values indicate thermalized, high density gas (the
larger the ratios the higher the temperature), while in galaxies with
CO(3-2)/CO(1-0) $<$ CO(2-1)/CO(1-0) lower density gas is indicated.

The galaxies with low to moderate CO(3-2)/CO(1-0) and CO(2-1)/CO(1-0)
values (He 2-10, Haro 2 and Mrk 86) are best fit by molecular gas
that is dense ($n_{H_{2}}~>10^{4}~ cm^{-3}$) and cold (T$_{K}~>$ 5 -
10 K).  The galaxies with high CO(3-2)/CO(1-0) and CO(2-1)/CO(1-0)
values (NGC 1569, NGC 3077 and NGC 5253) have a large range of
parameter space that fit the observed values.  There is a warm, low
density solution (T$_{K}~>$ 40 K; $n_{H_{2}}~\sim 10^{2.5}~ cm^{-3}$;
NGC 5253) or a warm, high density solution ($n_{H_{2}}~> 10^{3.5}~
cm^{-3}$; NGC 1569, NGC 3077).  For the remaining galaxies, Haro 3 and
II Zw 40, no constraints can be made (see \S6 for details).
``Point-source'' ratios preferentially bias the ratios downward,
ie. towards cooler temperatures or lower densities, so these derived
values reflect lower limits to the density and temperature if the true
source sizes are larger than $\sim 10^{''}$.

Table \ref{tbl4} summarizes the excitation conditions for the eight
galaxies.  The derived temperatures and densities of the molecular
clouds in these dwarf starburst galaxies are different from those
found in the Galactic disk and non-starburst dwarf galaxies, but
similar to those found in other, higher metallicity, starburst nuclei
\citep[eg.][]{YS91, BC92, WHEGGJRS92, DYSNY94, ABBJ95, PW98}.
\citet{SSLH92} come to a similar conclusion based on CO(2-1).

Despite the low metallicity, strong radiation fields and weak CO, CO
appears not to have become optically thin in the low J transitions of
dwarf starbursts.  A likely explanation for this is that if the CO
column density is high enough to shield itself against
photodissociation and hence maintain a detectable amount of CO
\citep[N$_{CO}$~\mbox{$\stackrel{<}{_{\sim}}$}$~10^{17}~ cm^{-2}$;][]
{LLPRBR94}, then that column density is high enough for the low J
transitions of CO to be optically thick \citep{PJVJB98}.  Over the
region of the starburst where CO can survive, CO might be expected to
be optically thick and resemble the physical conditions of higher
metallicity starbursts, with the difference between the two
environments (the weakness of CO in the low metallicity systems)
reflecting, predominately, increased beam dilution due to smaller ``CO
cloud size'' \citep[eg.][]{MB88, LLPRBR94, MW97}.

The large optical depths, the relatively high densities derived from
LVG modeling plus the small molecular cloud filling factor estimates
indicate that the CO cloud size may be smaller in these dwarf
starbursts than in spiral starbursts.  A similar effect is observed
for the LMC and SMC \citep*{IDVD86, RLB93, MET94}.  However, since
none of the molecular clouds are resolved here, we cannot say whether
this is due to photodissociated CO or is just a reflection of the gas
``traced by CO(3-2)'' with its larger critical density.

\subsection{Distribution of Warm, Dense Molecular Gas in Dwarf Starbursts}

Since bright CO(3-2) emission requires warmer temperatures and higher
densities than CO(1-0), it is expected that CO(3-2)/CO(1-0) would peak
on regions of active star formation.  There has been some evidence for
this in nearby spirals \citep[eg.]{GGC93, WWT97}, as well as some of the
nearby, less active dwarfs \citep{PW98,MHWS99}.

For the two galaxies in the sample with spatial information, He 2-10
and NGC 5253, we investigate the spatial variation in the
CO(3-2)/CO(1-0) ratio. In these two galaxies, CO(3-2)/CO(1-0) peaks
toward the starburst (Table \ref{tbl5}).  In the case of the fully
mapped galaxy, He 2-10, the CO(3-2) emission is more strongly peaked
towards the starburst than is CO(1-0), and the tidal features seen in
CO(1-0) (peak C) are less pronounced or ``missing'' all together .
\citet{BIK94} also find indications that the CO(1-0) source size is
slightly larger than the CO(3-2).  Likewise for NGC 5253, the
CO(2-1)/CO(1-0) value obtained from interferometric data indicates
that the ratios along the dust lane peak closest to, although not at
the starburst \citep{MT98}.  The CO(3-2) observations
demonstrate a similar trend.  While the three offset positions are not
deep enough to detect CO(3-2), the limits show that CO(3-2) is weaker
relative to CO(1-0) further from the starburst.

To further test the result that CO(3-2) is stronger in the central
starburst regions of the sample, we compared the observed line ratios
with the physical size covered on the galaxy.  The findings are
plotted in Figure 5. Note that the beamsizes of He 2-10 and NGC 5253
for the measured line ratios are larger than 22$^{''}$ and reflect the
CO(1-0) beam size of 40$^{''}$ and 44$^{''}$, respectively.  There is
a weak trend (correlation coefficient of r = -0.83) that the line
ratio decreases as the distance of the starburst galaxy increases,
which implies that including gas in the beam from farther out in the
galaxy lowers the ratio.  (Haro 3 appears to be somewhat of an
exception, but the line ratio of this galaxy appears ``contaminated''
by an optically thin velocity component [\S 6].)  This is consistent
with the notion that the highest temperatures and densities are
concentrated towards the starburst, with lower temperature and density
gas over the rest of the galaxy.  We conclude that this is a general
feature among dwarf starbursts, and the properties derived using the
higher J transitions of CO probably better represent the region
directly associated with the star formation.

\section{Masses, Column Densities and Conversion Factors}

Estimating molecular gas masses in dwarf galaxies can be a
particularly tricky business.  Dwarf galaxies seem to have a higher
standard conversion factor ($X_{CO}$, where $X=N(H_{2})/I_{CO}$) than
do large metal-rich spirals \citep{I86, MB88, E90, VH95, W95, AST96}.
Several methods are used to obtain molecular mass estimates.  One is
to assume that the molecular clouds are in virial equilibrium, and
that the linewidth is representative of the cloud mass.  However, with
the large beamsizes of single-dish observations, galaxy rotation
contributes significantly to the linewidth.  In fact, it is more
likely that the mass measured in this fashion represents an estimate
of the {\it dynamical} mass over the beam.  A second method for
estimating molecular mass is to assume CO is optically thin and
``count molecules'', using \citep[including He; eg.][page 191]{RW96}:
$$
M_{thin} = 173~(M_{\odot})~\left({8.5\times10^{-5}}\over{[CO/H_{2}]}
\right)
$$
$$
\times\left({e^{{33.2\over T_{ex}}}\over e^{{16.6\over T_{ex}}}-1}
\right)D^{2}_{mpc}~{\int T_{mb}dv \over K~km~s^{-1}}
$$
where the value is scaled relative to the Galactic [CO/H$_{2}$]
abundance ratio of $8.5 \times 10^{-5}$ \citep*{FLW82}.  This method
underestimates the total molecular mass, both because the CO emission
is optically thick (\S3.1), and the CO abundances, [CO/H$_{2}$], are
probably lower than the Galactic value for most of the dwarf
galaxies. Thirdly, molecular masses can be estimated using a
metallicity scaled conversion factor.  We use \citep{WSFMS88, PW98}:
$$
M_{mol}~=~1.23\times 10^{4}~ (M_{\odot}) \left({X_{CO}} \over 
{X_{CO_{gal}}}\right)
$$
$$
\times
\left(115 ~GHz \over {\nu} \right)^{2}~D^{2}_{mpc}~{S_{CO}\over R},
$$
where $X_{CO_{gal}}$ = $2.3\times 10^{20}~cm^{-2}(K km s^{-1})^{-1}$
\citep{SET88}, $S_{CO}$ is the CO(3-2) flux in Jy km s$^{-1}$
(where a conversion factor of 47.3 Jy/K is assumed), and R is the
CO(3-2)/CO(1-0) line ratio.  The scaling for the conversion factor is
quite uncertain due to the few data points, but we estimate it based
on the relationship derived from single-dish data \citep{AST96}.

Collected in Table \ref{tbl6} are the masses derived for each galaxy
by the three different methods.  We also compare the derived masses
with the dust mass estimated from the 100 $\mu m$ IRAS fluxes
\citep{TT86, MI94}.  The virial theorem gives a much larger mass than
any of the other methods, consistent with it tracing dynamical mass.
It should give a firm upper limit to the molecular mass present.  The
mass estimates obtained by assuming the CO(3-2) emission is optically
thin are more than 100 times lower than what is estimated from the
conversion factor method.  The M$_{thin}$ estimate gives a firm lower
limit to the amount of molecular gas.  The conversion factor method is
intermediate between the other two.  Therefore, even with its large
uncertainty, the best estimate is probably still the conversion factor
based estimate.  Since none of the molecular clouds are resolved
spatially or in velocity, a detailed investigation of the $X_{CO}$
must await higher resolution.  The CO(3-2) data does, however, appear
consistent with the I$_{CO}$ vs. metallicity relationship found from
CO(1-0) for dwarf and spiral galaxies \citep[Figure 3][]{VH95, W95,
AST96}.

\section{CO and Dust: Preferential CO Depletion?}

A correlation between CO and IR luminosity for large starbursts has
been known for some time \citep[see the][review and there references
therein]{YS91}.  In Figure 4, the infrared luminosity of this sample
plus the \citet{MHWS99} sample of non-dwarf nuclei are plotted versus
their CO(3-2) luminosity.  The CO(3-2) data for the dwarf starbursts
also show a tight correlation between L$_{CO32}$ and L$_{IR}$, with a
relation of:
$$
log(L_{IR})=(6.31\pm0.5) + (0.46\pm0.08)log(L_{CO32})
$$
(correlation coefficient of r=-0.94).  This can be compared to the
correlation we calculate using the \citet{MHWS99} dataset:
$$
log(L_{IR})=3.20 + 0.93log(L_{CO32})
$$
In dwarf starbursts, CO emission appears weaker relative to the
infrared emission than in the high metallicity galaxies.  The slope of
this relationship is substantially steeper than what is found from
nearby, high metallicity, starburst spirals.  The most metal-rich
galaxies in this sample follow the relationship found for non-dwarfs,
but the lower metallicity sources are weaker in L$_{CO32}$ relative to
dust than would be extrapolated from non-dwarfs.  While the sample of
dwarf galaxies is too small to prove this fact conclusively, it
indicates that there may be a break in the slope of L$_{CO32}$
vs. L$_{IR}$ at the faint, low metallicity end.  \citet{TKS98} also
suggest a rapid fall off in L$_{CO}$ at metallicities of about 1/10
Z$_{\odot}$ based on a different line of reasoning.  Further
observations are key to investigating the validity of this trend.

The weakening of L$_{CO}$ relative to L$_{IR}$ is not necessarily
unexpected because of the strong radiation field and low metallicity
of these dwarf starbursts. These simultaneously lead to higher dust
temperatures and hence L$_{IR}$ and to increased CO photodissociation.
Dust survives longer than CO in strong radiation fields, so CO should
disappear faster than dust.
\begin{figure}
\centerline{\psfig{figure=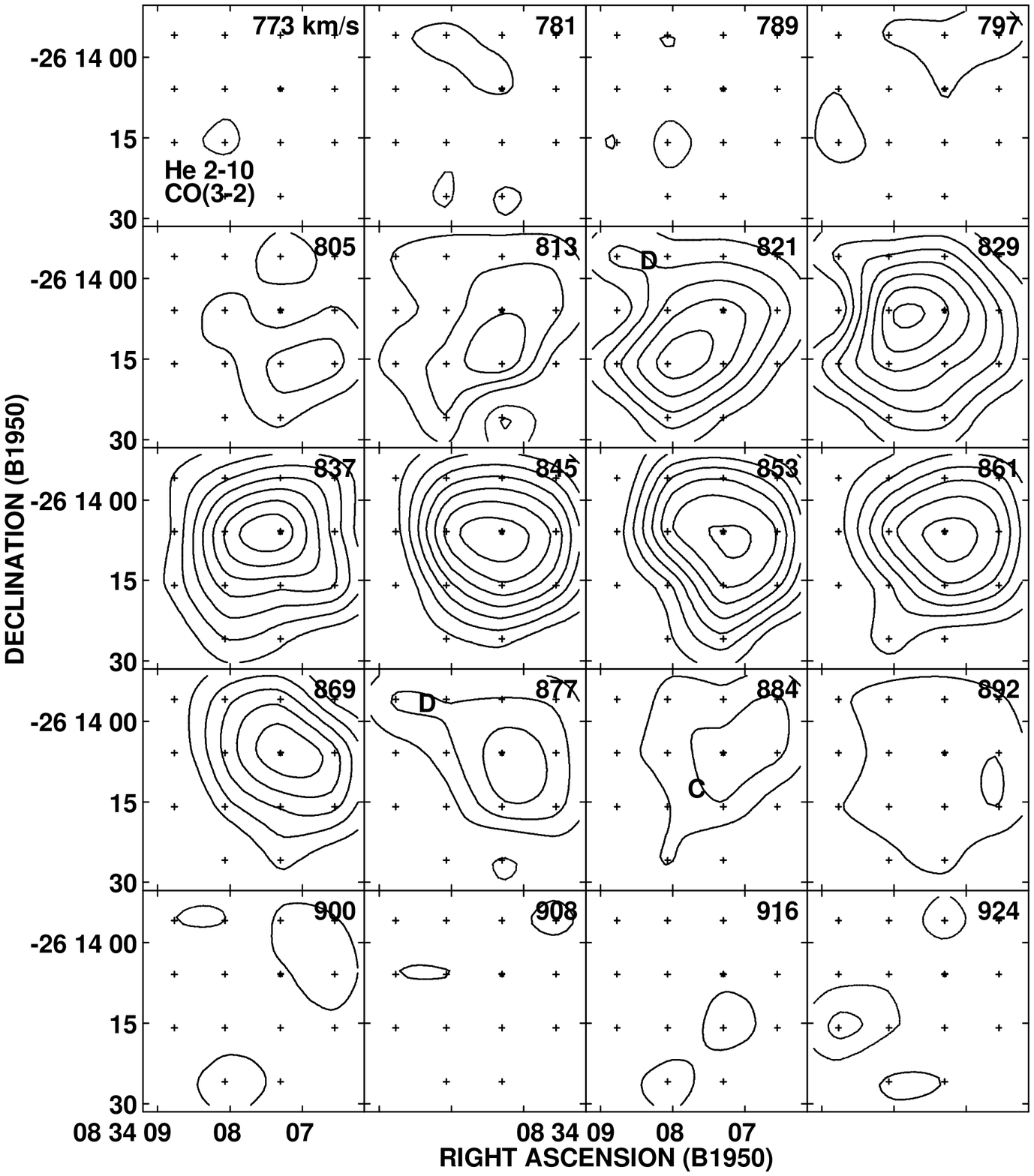,width=4.5cm,angle=00}}
\caption{The channel maps of He 2-10.  Panels are
displayed in intervals of 8 km s$^{-1}$.  The V$_{LSR}$ of each
channel is displayed in the upper right hand corner of each
frame. Contours are 1, 2 ..., 8 times the 2$\sigma$ temperature of
0.035 K.  The asterisk marks the location of the optical starburst and
center position (0,0).  Each additional pointing is marked by a cross.
The feature denoted by Kobulnicky et al. (1995) as ``C'', is also
marked at its correct location and velocity, and the location and the
two velocities contributing to the ``detached'' feature is marked with
a ``D'' (Kobulnicky et al. 1995).}
\end{figure}

\begin{figure}
\centerline{\psfig{figure=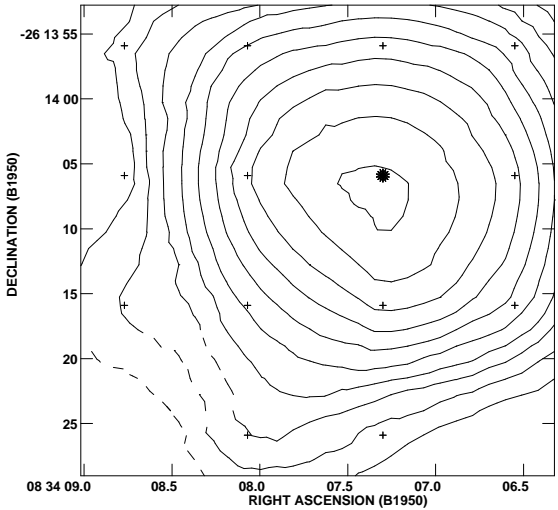,width=7.5cm,angle=00}}
\caption{The integrated intensity map of He
2-10. The integrated intensity map was mapped from only the channels
displayed in Figure 2, (773 to 924 km s$^{-1}$).  Emission below
1.3$\sigma$ in each channel map were cut.  Contours are 1, 2 ..., 10
times the 2$\sigma$ value of 1.6 K km s$^{-1}$.  Again, the
asterisk marks the location of the optical starburst and center
pointing (0,0) and each additional pointing is marked by a cross.  The
map is uncertain at the (20$^{''}$,-20$^{''}$) position because there
is no observation there.  To represent this fact the contours are
dashed.}
\end{figure}

\begin{figure}
\centerline{\psfig{figure=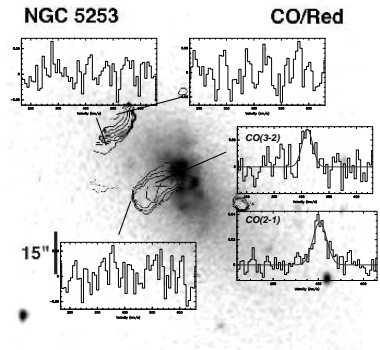,width=7.5cm,angle=00}}
\caption{The spectra of NGC 5253 overlaid on the
OVRO millimeter interferometer CO(1-0)/Red image of Turner, Beck \&
Hurt (1997) (their Figure 2).  Each observation is displayed with a
line running from the spectrum to the location at which it was
obtained.  The vertical axis is main-beam temperature and ranges from
-0.065 to 0.065 K.  The horizontal axis is velocity and ranges from
180 to 620 km s$^{-1}$.  The spectra are hanning smoothed to a
velocity resolution of 8 km s$^{-1}$.  For the central position the
new CO(2-1) spectrum is included underneath the CO(3-2) spectra.  For
the detections the best fit gaussian is displayed.}
\end{figure}

\section{Notes on the Individual Galaxies}

{\it He 2-10:} In Figure 6, the channel maps of He 2-10 are presented.
A 14-point map of He 2-10 was made on a 10$^{''}$ grid in RA and
Dec. around the brighter optical starburst component \citep{VC92}.
The ($\Delta \alpha,\Delta \delta$) = 0,0 position of He 2-10 was
reobserved on February 25 to confirm its line strength.  Pointings
were extended towards the southeastern portion of the galaxy where a
bar-like tidal feature is seen in the CO(1-0) interferometer map
\citep{KDSHC95}.  We detect emission in every position except
(20$^{''}$,0).  The peak T$_{mb}$ occurs at the central position and
has a value of 0.27 K.  This value is about 80\% of the value
\citet{BIK94} find for CO(3-2) after we convolve to the same beamsize.
The CO(3-2) emission at this position has a V$_{FWHM}~\sim$ 57 km
s$^{-1}$ centered at 849 km s$^{-1}$, with detectable emission
spanning a velocity range of 810 - 890 km s$^{-1}$ (LSR).

We do not detect any obvious emission to a level of 40 mK associated
with the kinematically distinct feature labeled by \citet{KDSHC95}
with the letter ``C''.  There is some hint of a southeastern extension
seen at 853 km s$^{-1}$ and 861 km s$^{-1}$, but this is separated by
better than two channels (16 km s$^{-1}$) from their velocity and is
not likely to be the same source.  On the other hand, we do detect
emission that matches both in position and velocity space with the
weak detached feature seen in the \citet{KDSHC95} interferometric
CO(1-0) map.  \citet{KDSHC95} interpret this detached feature
questionably as noise, but since we see indications of emission in
CO(3-2) at the same location and velocities, this feature may be real.
This feature is weak, with a T$_{MB}~\sim$ 0.08 K, and centered at
$\alpha~=~8^{h}34^{m}08.^{s}65$ and $\delta~=~-26^{o}13'56.^{''}5$.

Figure 7 shows the integrated CO(3-2) intensity map of the nuclear
region of He 2-10.  The emission from the central peak is more
dominant than in  CO(1-0).  Within pointing uncertainties, the peak
of the integrated intensity coincides with the position of the
brighter optical starburst component.  Of the three extensions seen,
one extends to the northeast along a p.a. $\simeq 50^{o}$, while the
other two extend to the southeast.  We interpret the two extensions to
the SE as one feature along a p.a. of 130$^{o}$ and attribute the
apparent split into two features to the lack of a pointing at
(20$^{''}$,-20$^{''}$).  This extension matches what is observed in
the interferometric CO(1-0) and CO(2-1) maps and is near the site of
an offset star cluster \citep{KDSHC95, BK98, MT98}.

The centrally peaked component has a FWHM of 26$^{''}$.  When
deconvolved from the beam, we estimate that the source size is
13$^{''}\pm 3^{''}$ (560$\pm$110 pc at 9 Mpc).  Since the
observations of \citet{BIK94} are at a different resolution than ours,
we can compare the observed main-beam temperatures to get a second
estimate of the source size.  Again, 13$^{''}$ is obtained.  Because
both methods agree, and are consistent with the size estimated from
the high resolution interferometer CO(2-1) map \citep{MT98},
we conclude that this is a good representation of the CO(3-2) source
size.

Assuming a source size of 13$^{''}$ for He 2-10, we obtain
CO(3-2)/CO(2-1), CO(3-2)/CO(1-0), and CO(2-1)/CO(1-0) line ratios of
1.0$\pm$0.2, 0.61$\pm$0.1 and 0.59$\pm$0.1 respectively.  These are
slightly lower than the previous results of \citet{BIK94}.
CO(3-2)/CO(2-1) $>$ 1.0 combined with a high $^{12}$CO/$^{13}$CO
isotopic ratio led them to conclude that the CO emission is optically
thin.  With the advantage of a fully sampled map and CO(3-2)
observations with beam matched to their J=2-1 observations, these
observations do not necessarily require optically thin gas. LVG
modeling yields a best fit solution of $n_{H_{2}}~> 10^{3.5}~ cm^{-3}$
and T$_{k}~\sim$ 5 - 10 K, however the $\chi^{2}$ on the solution is
high.  These gas temperatures are slightly cooler than the $\sim$20 K
estimated from LTE modeling.  The $^{12}$CO(1-0)/$^{13}$CO(1-0) line
ratio was also modeled.  A model with [$^{12}$CO/$^{13}$CO] = 40 and
$dv/dr~=~ 1~ km~ s^{-1}pc^{-1}$ and a $^{13}$CO depleted model with
[$^{12}$CO/$^{13}$CO] = 120 were run (Figure 2).  Both CO abundance
models are inconsistent with the solution found from the $^{12}$CO
transitions.  The $^{13}$CO line in He 2-10 is too weak
($^{12}$CO(1-0)/$^{13}$CO(1-0) \mbox{$\stackrel{<}{_{\sim}}$} 22)
relative to what is predicted from the $^{12}$CO lines, as
\citet{BIK94} noted.  In order to have $^{12}$CO(1-0)/$^{13}$CO(1-0)
$\sim$ 22, the $^{12}$CO lines must be optically thin, unless the
abundance is {\it severely} depleted.  However nowhere over the
CO(3-2) solution space is the modeled $^{12}$CO optically thin.
Therefore, given the high $\chi^{2}$ on the $^{12}$CO solution and
that the $^{13}$CO observations are not consistent with the simple LVG
modeling, the solution should be considered uncertain. $^{13}$CO is
probably preferentially photodissociated or that source sizes are much
smaller in $^{13}$CO than $^{12}$CO.  As far as the high isotopic
ratios obtained are concerned \citep{BIK94, KDSHC95}, the optically
thinner $^{13}$CO lines are less capable of self-shielding than
$^{12}$CO, and are likely to be preferentially photo-dissociated by
the intense radiation fields.  So the abundance and spatial extent of
$^{13}$CO may be greatly diminished with respect to $^{12}$CO
\citep*[eg.][]{VB88, LLPRBR94, WBV96}.

{\it NGC 5253:} The four pointings for NGC 5253 were chosen based on
the CO(1-0) interferometric map made at Owens Valley Millimeter
Observatory (OVRO) \citep*{TBH97}.  We observed two pointings
associated with the central dust lane, and two toward a possible
molecular cloud at the edge of the OVRO primary beam.  In Figure 8,
the four CO(3-2) and one CO(2-1) spectra are overlaid on the OVRO
CO(1-0) map \citep[Figure 2 of][]{TBH97}.  The measured CO(3-2)
intensity is about 5 - 6 $\sigma$, and both the velocity centroid
(V$_{o} \sim 420$ km s$^{-1}$) and the line width ($\Delta V_{FWHM}
\sim 45$ km s$^{-1}$) match published single-dish CO(1-0) detections
\citep{WH89, TKS98}.  For the three other spectra there are no
detections to 3$\sigma$.  LVG modeling yields densities of
$n_{H_{2}}~> 10^{2.5}~ cm^{-3}$, and T$_{k}~>$ 40 K, with the best fit
solutions obtained for the lower densities.

{\it NGC 1569:} We detect CO(3-2) in this galaxy with I$_{CO}$=2.2 K
km s$^{-1}$ (T$_{MB}$ = 63 mK) towards the HII region C \citep{W91}.
This corresponds to the central position observed by \citet{GBJM96} in
CO(1-0) and CO(2-1).  The line width of CO(3-2) is slightly larger
(32.3 km s$^{-1}$) than that of the CO(1-0) observations (23.6 km
s$^{-1}$).  The line center is shifted to slightly higher velocities
than the CO(1-0) (-69 relative to -81 km s$^{-1}$ for CO(1-0)).  The
CO(2-1) line center is intermediate between the two, 78.4 km s$^{-1}$.
Therefore, while the velocity centroids are roughly consistent, there
is a marginal trend that the emission from the higher J transitions
tend towards increasing velocity.  Comparing these higher velocities
with the velocity field obtained from HI \citep{R80, SI98} indicates
that CO(3-2) peaks slightly closer to the super-star cluster, A, than
does CO(1-0), and is probably associated with GMC 2 of \citet{THKG99}.

Interestingly, there is a secondary peak seen at -140 km s$^{-1}$.
This peak is at the 4$\sigma$ level, and the line profile is similar
to the main peak.  HI is also present at this velocity, giving more
evidence this feature may be real.  We have inspected the CO(1-0) and
CO(2-1) spectra of \citet{YGH84}, \citet{TKS98} and \citet{GBJM96} for
similar features.  Unfortunately, the spectra of \citet{GBJM96} does
not extend that far in velocity, and while the \citet{YGH84} spectrum
shows indications of a feature at this velocity, it does not seem to
be confirmed by the more sensitive observations of \citet{TKS98}.
Therefore, until further observations can be obtained, we assume it is
not real.  Since both the CO(3-2)/CO(1-0) and the CO(2-1)/CO(1-0) line
ratios are $\simeq ~ 1$, a large region of parameter space is
acceptable to the LVG models.  The best fits represent an arch running
from T$_{K}~\sim$ 100 K and $n_{H_{2}}~\simeq 10^{4.5}~ cm^{-3}$ down
to T$_{K}~\sim$ 40 K and $n_{H_{2}}~\simeq 10^{3.5}~ cm^{-3}$ and back
over to T$_{K}~\sim$ 100 K and $n_{H_{2}}~\sim 10^{2.4}~ cm^{-3}$.

{\it NGC 3077:} Observations were centered on the location of the
interferometer CO(2-1) peak \citep{MT98}.  This is equivalent to the
(-10,0) position of \citet{BSH89}.  The CO(3-2) line width of 51 km
$^{-1}$ is slightly wider than observed by \citet{BSH89} in CO(1-0)
(34 km s$^{-1}$), while the centroid is consistent with what is
observed in the lower transitions.  Based on the CO(2-1)/CO(1-0) line
ratio discussed in \citet{BSH89}, we estimate a CO(3-2)/CO(1-0) line
ratio of 1.1.  \citet{BSH89} obtain a CO(2-1)/CO(1-0) line ratio of
0.82.  For NGC 3077, the best fit LVG solution corresponds to
T$_{k}~\sim$ 30 K and $n_{H_{2}}~> 10^{3.5}~ cm^{-3}$, with an
acceptable range of solutions covering T$_{K}~> 20$ K and $n_{H_{2}}~>
10^{3}~ cm^{-3}$.

{\it Haro 2:} We observed CO(3-2) at the same position as CO(1-0) and
CO(2-1) of \citet{SSLH92}.  The fitted line width is 69 km s$^{-1}$,
but the line appears distinctly non-gaussian.  The SNR of the spectrum
is to low to warrant any multi-component fit, but the spectrum is
consistent with two roughly equal intensity gaussians separated by
about 40 km s$^{-1}$.  For Haro 2, the best fit solution obtained from
LVG modeling is $n_{H_{2}}~> 10^{4}~ cm^{-3}$ and T$_{k}~\sim$ 5 K,
but the $\chi^{2}$ are high, similar to what is found for He2-10.
However, the low ratios obtained for both transition require that 
T$_{k}$ be low and $n_{H_{2}}$ be high.

{\it Haro 3 (NGC 3353):} CO(3-2) observations are centered on the
starburst region B \citep{SSMRM96}.  The fitted gaussian yields a line
width of 53 km s$^{-1}$ which is slightly narrower than CO(1-0)
\citep[$\Delta v_{10}$ = 64 km s$^{-1}$;][]{SSLH92}.  The CO(3-2)
spectrum shows a narrow brighter component at 960 km s$^{-1}$
($v_{LSR}$ = 950 km s$^{-1}$).  The same peak is seen in both the
CO(1-0) and CO(2-1) spectra of \citet{SSLH92}, and matches the
velocity of H$\alpha$ at the starburst region, B1, of \citet{SSMRM96}.
Therefore, we conclude this feature is real. Inspecting the spectra of
\citet{SSLH92} a main-beam temperature ratio of 1.3 is estimated for
this component.  The large CO(3-2)/CO(1-0) line ratio is inconsistent
with any region of parameter space indicated by the CO(2-1)/CO(1-0)
line ratio, so no LVG solutions is found.  This component is
potentially optically thin and may be causing an overestimate of the
global CO(3-2)/CO(1-0) ratio in Haro 3.

{\it II Zw 40:} The beam was centered on the star-formation peak and
is within an arcsecond of the observed CO(2-1) and CO(1-0) position of
\citet{SSLH92}.  The line is not detected.  Our limit for the CO(3-2)
main-beam temperature is 7 mK.  There appears to be no hint of a line
at the $1\sigma$ level in either of the previously published CO(1-0)
velocity centroids, $\sim$ 770 km s$^{-1}$ \citep{SSLH92} or $\sim$
850 km s$^{-1}$ \citep{TY87}.  LVG modeling does not
provide a useful constraint on the gas properties because CO(3-2) is
not detected.  The low CO(3-2)/CO(1-0) ratio implies that the
molecular gas must not be very warm and dense.

{\it Mrk 86 (NGC 2537):} We do not detect any CO(3-2) emission to a
limit of 11 mK.  \citet{SSLH92} detected weak CO(1-0) and CO(2-1)
at a V$_{LSR}$ of 460 km s$^{-1}$.  There is a hint of a very weak
line at V$_{LSR}$ = 428 km s$^{-1}$, but only at 1.5$\sigma$.  The 
low CO(3-2)/CO(1-0) ratio limit and the low CO(2-1)/CO(1-0) ratio 
imply LVG fits quite similar to those of Haro 2.

\section{Conclusions}

We have observed CO(3-2) in a sample of eight dwarf galaxies with very
intense recent star formation.  The galaxies range in metallicity from
1 Z$_{\odot}$ to $\sim$0.1 Z$_{\odot}$.  CO(3-2) was detected in six
of the eight galaxies.  The galaxies are all quite weak (T$_{mb}~\sim~
50$ mK), with the exception of He 2-10, which was mapped.  The two
galaxies not detected (II Zw 40 and Mrk 86) are two of the lowest
metallicity galaxies.

The CO(3-2)/CO(1-0) line ratio ranges from 0.37 - 1.1 for the six
detected galaxies.  With the possible exception of one velocity
component in Haro 3, all the galaxies appear dominated by optically
thick CO emission, even though they have low metallicity and strong
radiation fields.  This is consistent with the fact that the only
place that CO can survive intense radiation fields is where it is
self-shielded, and thus optically thick.

The error weighted mean CO(3-2)/CO(1-0) line ratio is 0.60$\pm$0.06.
This is virtually identical to what is seen in high metallicity
starbursts.  While CO(3-2) is very weak on an absolute scale, it is as
bright relative to CO(1-0) as in high metallicity starbursts.  In
general, in terms of the physical properties of the molecular gas,
dwarf starbursts appear not to be particularly unusual examples of the
starburst family.  They appear to be similar to their high metallicity
counterparts, with the exception that CO has a smaller filling factor
and hence weaker intensities.

LVG models from the CO line ratios imply that for He 2-10, Haro 2
and Mrk 86, the molecular gas traced by CO is cool (T$_{k}~\sim$10 K)
and dense ($n_{H_{2}}$\mbox{$\stackrel{<}{_{\sim}}$} $10^{4}~
cm^{-3}$).  For NGC 5253, the molecular gas appears warm (T$_{k}~>~40$
K) and moderately dense ($n_{H_{2}}\sim 10^{2.5-3}~ cm^{-3}$). The
molecular gas in NGC 1569 and NGC 3077 are also warm but somewhat more
dense ($n_{H_{2}}~>~ 10^{4}~ cm^{-3}$).

For the two galaxies that have been observed with multiple pointings,
He 2-10 and NGC 5253, we find that the CO(3-2)/CO(1-0) line ratio
decreases away from the starburst.  The entire sample shows a trend
where the ratio is lower for the galaxies where the beam includes more
``non-starburst'' gas (ie. covers larger fractions of the galaxy).
This is interpreted as due to a drop in gas temperature and density
away from the starburst.

The combination of low filling factors derived from the CO(3-2)
emission, the fact that CO(3-2) systematically remains optically thick
in spite of low metallicities and strong radiation fields, and a trend
found between L$_{CO}$ and L$_{IR}$ implies the lower metallicity
galaxies are more strongly depleted in CO relative to dust than higher
metallicity galaxies.

\acknowledgements We gratefully thank Tom Phillips, Dominic Benford
and Atilla Kovacs for useful discussions and help in making the
observations.  We thank Gene Serabyn for obtaining the CO(2-1)
spectrum of NGC 5253.  We also thank Chip Kobulnicky for useful
discussions and providing us with the OVRO CO(1-0) data for He 2-10.
We also thank the referee, Dr. Christine Wilson for many helpful
comments.  This work was supported in part by NSF grant AST9417968 and
by the Sackler Center for Astronomy at Tel Aviv University.
This research has made use of the NASA/IPAC extragalactic database
(NED).

\begin{deluxetable}{lcccccc}
\tablenum{1}
\tablewidth{0pt}
\tablecaption{The Galaxy Sample} 
\tablehead{
\colhead{Galaxy} &\colhead{RA} &\colhead{Dist.}&
\colhead{IRAS 60/100$\mu$m} &\colhead{log(L$_{FIR}$)} & 
\colhead{log(M$_{HI}$)}&\colhead{References} \\
\colhead{} &\colhead{DEC} & \colhead{(Mpc)}& \colhead{(Jy)} 
&\colhead{(L$_{\odot}$)} & \colhead{(M$_{\odot}$)} & \colhead{} \\ 
\colhead{} &\colhead{} & \colhead{-M$_{B}$}&
\colhead{T$_{D}$\tablenotemark{a}}
&\colhead{M$_{D}$\tablenotemark{b}} & \colhead{[O/H]} & \colhead{}
\\
\colhead{} & \colhead{(B1950)}& \colhead{} &\colhead{(K)}
&\colhead{(M$_{\odot}$)} &\colhead{} & \colhead{}}
\startdata
He 2-10&08:34:07.3&9.0&24.0/26.4&9.43 &8.53& 1,9,10,11\\
&-26:14:05.9&17.4&43&5.47&8.93& \\
NGC 5253&13:37:05.8&4.1&30.9/29.0&8.98 &8.3&12,15,19,22,23\\
&-31:23:19.0&17.2&46&4.73&8.10& \\
NGC 1569&04:26:02.0&2.2&46.3/50.7&8.65 &7.93&5,6,7,13,22 \\
&64:44:31.0&16.9&43&4.52&8.19& \\
NGC 3077&09:59:20.0&3.25&14.7/26.9&8.70 &9.00&2,4,16,17,21 \\
&68:58:30.0&16.1&34&4.99&9.02& \\
Haro 2&10:29:22.7&20.3&4.7/5.3&9.60&8.68&1,14,18,22 \\
&54:39:24.0&18.4&42&5.51&8.4& \\
Haro 3&10:42:16.5&13.7&5.1/6.4&9.30&8.79&8,18,20,22 \\
&56:13:23.0&17.5&41&5.16&8.4& \\
II Zw 40&05:53:04.9&9.2&6.5/5.7&9.04&8.3&3,8,18,22 \\
&03:23:06.0&16.2&48&4.66&8.15& \\
Mrk 86&08:09:42.8&6.3&3.2/6.3&8.45&8.28&8,18,22 \\
&46:08:33.0&16.6&33&5.00&8.05\tablenotemark{c}& \\
\enddata
\tablenotetext{a}{using R(60/100)=0.6$^{-4}$(e$^{136/T_{D}}$-1)/
(e$^{226/T_{D}}$-1), (Thronson \& Telesco 1986)}
\tablenotetext{b}{using M$_{D}$=5D$^{2}$F$_{Jy}$(e$^{144/T_{D}}$-1),
(Thronson \& Telesco 1986)}
\tablenotetext{c}{Extrapolated from the absolute magnitude based on 
the relationship of Skillman, Kennicutt \& Hodge 1989}
\tablerefs{(1)Baas et al. 1994; (2)Cottrell 1976; (3)Garnett1990; 
(4)Heckman 1980; (5)Israel 1988; (6)Israel \& de Bruyn 1988; 
(7)Israel \& van Driel 1990; (8)Klein et al. 1991; (9)Kobulnicky et al. 1995; 
(10)Kobulnicky \& Johnston 1999; (11)Kobulnicky et al. 1998; (12)Kobulnicky 
\& Skillman 1995; (13)Kobulnicky \& Skillman 1997; (14) Kunth \& Jobert 1985; 
(15)Marconi et al. 1994; (16)Melisse \& Israel 1994; (17)Niklas et al. 1995; 
(18)Sage et al. 1992; (19)Sandage 1994; (20)Steele et al. 1996; 
(21)Tammann \& Sandage 1968; (22)Thronson \& Telesco 1986; (23)Turner et al. 
1998}
\label{tbl1}
\end{deluxetable}

\begin{deluxetable}{lcccccc}
\tablenum{2}
\tablewidth{0pt}
\tablecaption{Results}
\tablehead{
\colhead{Galaxy}&\colhead{Transition} & \colhead{Offset} &
\colhead{V$_{LSR}$} & \colhead{$\Delta$V} & 
\colhead{T$_{mb}$} &\colhead{I$_{co}$}\\
\colhead{} & &\colhead{($\Delta\alpha^{''}$,
$\Delta\delta^{''}$)} &\colhead{(km s$^{-1}$)} & \colhead{(km s$^{-1}$)} & 
\colhead{K} & \colhead{K km s$^{-1}$}}
\startdata
He 2-10&3-2&(0,0)&849$\pm$1.0& 56.6$\pm$2.3 & 0.27$\pm$0.018 &16.6$\pm$0.6 \\
&&(0,-10)&846$\pm$1.6 & 63.8$\pm$3.7 & 0.20$\pm$0.018 & 13.3$\pm$0.7\\
&&(-10,0) &854$\pm$1.5 & 58.8$\pm$3.6 & 0.19$\pm$0.02 &11.9$\pm$0.6 \\
&&(0,10) &850$\pm2.0$ & 63.4$\pm$4.8 &0.15$\pm$0.022 & 10.3$\pm$0.7 \\
&&(10,0) &846$\pm1.4$ &50.0$\pm$3.4 & 0.24$\pm$0.022 & 12.6$\pm$0.7 \\
&&(10,10)&849$\pm$3.0 & 61.7$\pm7.0$ &0.12$\pm$0.016 & 7.77$\pm$0.8 \\
&&(10,-10) &833$\pm$1.8 & 37.2$\pm4.4$ &0.19$\pm$0.023 &7.47$\pm$0.7 \\
&&(10,-20) &872$\pm$14.4 & 140$\pm$25 &0.047$\pm$0.024 & 7.00$\pm$1.4 \\
&&(20,10) &826$\pm$3.3 &22.5$\pm$7.1 & 0.090$\pm$0.022 & 2.16$\pm$0.6 \\
&&(0,-20)&845$\pm$5.6 & 33.0$\pm$4.4 & 0.072$\pm$0.030 & 2.53$\pm$0.8 \\
&&(-10,-10) &851$\pm$2.3 & 59.7$\pm$6.5 & 0.142$\pm$0.020 &9.06$\pm$0.8 \\
&&(20,0) &\nodata &\nodata & $\leq$0.053 & \nodata \\
&&(-10,10) &855$\pm$3.0 & 73.7$\pm$7.7 & 0.094$\pm$0.019 &7.4$\pm$0.6 \\
&&(20,-10) &825$\pm$3.9 & 26.7$\pm$10.8 & 0.103$\pm$0.022 & 2.91$\pm$0.7 \\
&&&&&& \\
NGC 5253&3-2&(0,0)&421$\pm$4.2 & 45.4$\pm$8.7 & 0.070$\pm$0.021 & 
3.39$\pm$0.6\\
&&(10,-10) &\nodata &\nodata&$\leq$0.073 &\nodata \\
&&(20,20)&\nodata &\nodata &$\leq$0.069 &\nodata \\
&&(15,25) &\nodata&\nodata &$\leq$0.075 &\nodata \\
&2-1&(0,0) &407$\pm$2.9 &71.6$\pm$9.0 & 0.034$\pm$0.006 & 
2.58$\pm$0.3 \\
NGC 1569 &3-2&(0,0) &-68.4$\pm$2.0
&32.3$\pm$4.0 &0.063$\pm$0.01 &2.2$\pm$0.3\\
&&&&&& \\
NGC 3077 &3-2&(0,0) & 7.80$\pm$2.9&50.9$\pm$7.0 &0.10$\pm$0.02 &
5.5$\pm$0.6\\
&&&&&& \\
Haro 2 &3-2&(0,0) &1422$\pm$4.6
&68.8$\pm$9.8 &0.032$\pm$0.01 &2.3$\pm$0.3 \\
&&&&& \\
Haro 3&3-2&(0,0) &951$\pm$4.6 &52.6$\pm$14 &0.034$\pm$0.01 &1.9$\pm$0.4
\\
&&&&&& \\
II Zw 40&3-2&(0,0) &\nodata &\nodata &$\leq$0.021
&$\leq$0.9 \\
&&&&&& \\
Mrk 86 &3-2&(0,0) &\nodata &\nodata&$\leq$0.033 &$\leq$1.8 \\
\enddata
\tablecomments{Offsets are in arcseconds from the central position
listed in Table 1.  Errors are 1$\sigma$ from a gaussian fit, based
only on the uncertainty in the spectrum.  The noise in the gaussian
fit for Haro 2 is worse than the actual noise is the spectrum,
implying it is not gaussian shaped (see text).  For non-detections
3$\sigma$ limits are quoted. The uncertainties in absolute calibration
were not included.}  
\label{tbl2}
\end{deluxetable}

\begin{deluxetable}{lcccc}
\tablenum{3}
\tablewidth{0pt}
\tablecaption{CO Line Ratios}
\tablehead{
\colhead{Galaxy} &\colhead{Method} &\colhead{3-2/1-0} 
&\colhead{3-2/2-1} &\colhead{2-1/1-0}}
\startdata
He 2-10&point source (PS)&0.50$\pm$0.1&0.87$\pm$0.2&0.48$\pm$0.07 \\
(0,0)&uniform filling (U)&1.7$\pm$0.3&0.96$\pm$0.2 &1.73$\pm$0.3 \\
&deconvolved (D)&$^{*}$0.61$\pm$0.1&0.96$\pm$0.2&$^{*}$0.59$\pm$0.1 \\
NGC 5253&PS&0.65$\pm$0.2&0.70$\pm$0.2&0.93$\pm$0.2 \\
(0,0)&U&2.6$\pm$0.7&1.3$\pm$0.4&2.0$\pm$0.4 \\
&D&$^{*}$0.70$\pm$0.2&0.73$\pm$0.2&$^{*}$0.96$\pm$0.25 \\
NGC 1569&PS&$^{*}$1.00$\pm$0.2&2.5$\pm$0.6&0.42$\pm$0.1 \\
&U&1.10$\pm$0.2&0.88$\pm$0.2&1.10$\pm$0.2 \\
&D&\nodata&\nodata&$^{*}$1.10$\pm$0.2 \\
NGC 3077&PS&$^{*}$0.95$\pm$0.2&2.1$\pm$0.7&0.55$\pm$0.1 \\
&U&1.04$\pm$0.25 &0.83$\pm$0.2&1.2$\pm$0.3 \\
&D&\nodata&\nodata&$^{*}$0.82$\pm$0.2 \\
Haro 2&PS&$^{*}$0.34$\pm$0.1&1.1$\pm$0.2& $^{*}$0.31$\pm$0.10\\
&U&0.37$\pm$0.1&0.96$\pm$0.2&1.0$\pm$0.1 \\
Haro 3&PS&$^{*}$0.89$\pm$0.3&1.7$\pm$0.5&$^{*}$0.50$\pm$0.09 \\
&U&0.97$\pm$0.3&0.57$\pm$0.2& 1.7$\pm$0.3\\
II Zw 40&PS&$\le$0.39&\nodata&0.58$\pm$0.16 \\
&U&$\le$0.43&$\le$0.24&1.78$\pm$0.32  \\
Mrk 86&PS&$\le$0.41&\nodata&0.31$\pm$0.11 \\
&U&$\le$0.45&$\le$0.52 &1.0$\pm$0.36 \\
\enddata
\tablecomments{See the text for the description of the method for
finding each ratio and the estimation of the errors.  The slight
difference between the IRAM 30m (1-0) beam and the CSO (3-2) beam has
been accounted for in the calculations of the ratios.  An asterisk
denotes which value of the line ratio is used in the LVG modeling.
All line ratios are based on a 22$^{''}$ beamsize except He 2-10 and
NGC 5253, which are based on a 40$^{''}$ beamsize. The data for the
2-1 \& 1-0 emission comes from: He 2-10: Baas et al. 1994; NGC 5253;
Wiklind \& Henkel 1989 and this paper; NGC 1569: Greve et al. 1996;
NGC 3077: Becker et al. 1989; the rest: Sage et al. 1992.}
\label{tbl3}
\end{deluxetable}

\begin{deluxetable}{lcccccccc}
\tablenum{4}
\tablewidth{0pt}
\tablecaption{Molecular Gas Properties}
\tablehead{
\colhead{} & \colhead{He 2-10} &\colhead{NGC5253} & \colhead{NGC1569}
& \colhead{NGC3077}&\colhead{Haro2} & \colhead{Haro3} &\colhead{IIZw40} 
&\colhead{Mrk86} 
}
\startdata
LTE:&&&&&&&\\
T$_{d}$&43&46&43&34&42&41&48&33\\
T$_{ex}$&11&15&$>$50&$>$50&6&$>$50&$<$7&$<$7\\
f$_{a}$&0.02&0.005&$<$0.001&$<$0.002&0.005&$<$0.001&$<$0.004&$<$0.005\\
$<G_{o}>$\tablenotemark{a}&$10^{3.4}$&$10^{3}$&$10^{2.5}$&$10^{2.7}$
&$10^{3.6}$&$10^{3.3}$&$10^{3}$&$10^{2.4}$\\
LVG:&&&&&&&& \\
T$_{k}$&10\tablenotemark{b}&30&$>$40&\mbox{$\stackrel{>}{_{\sim}}$}20&5
&\nodata&\nodata&5\\
n$_{H_{2}}$&\mbox{$\stackrel{>}{_{\sim}}$}$10^{3.5}$\tablenotemark{b}
&$\sim10^{3}$&
$>10^{2.5}$&
\mbox{$\stackrel{>}{_{\sim}}$}$10^{3}$&
\mbox{$\stackrel{>}{_{\sim}}$}$10^{4}$&\nodata&\nodata&
\mbox{$\stackrel{>}{_{\sim}}$}$10^{4}$\\
$\sqrt{n}\over T_{k}$&\mbox{$\stackrel{>}{_{\sim}}$}5.6\tablenotemark{b}
&1.0&
$\sim$1.5&\mbox{$\stackrel{>}{_{\sim}}$}1.5&20&\nodata&\nodata&20\\
&&&&&&&&\\
n$_{H_{2}}T_{k}$
&\mbox{$\stackrel{>}{_{\sim}}$}$3\times10^{4}$\tablenotemark{b}&
$3\times 10^{4}$&\mbox{$\stackrel{>}{_{\sim}}$}$1\times10^{5}$&
\mbox{$\stackrel{>}{_{\sim}}$}$ 2\times10^{4}$
&\mbox{$\stackrel{>}{_{\sim}}$}$5\times10^{4}$
&\nodata&\nodata&
\mbox{$\stackrel{>}{_{\sim}}$}$5\times 10^{4}$\\
\enddata
\tablenotetext{a}{Using the ``OB association'' relationship,
$<G_{o}>\simeq 1.2\times 10^{-6}$ L$_{IR}$ of Wolfire, Tielens \&
Hollenbach 1990, with R = $\lambda ~\sim$ 100 pc which is typical of 
galactic nuclei.  This may be an underestimate of $\lambda$ and thus 
of $<G_{o}>$, since these dwarfs are dust poor and are potentially 
more ``porous'' to radiation than metal rich galaxies (Petitpas 
\& Wilson 1998).}
\tablenotetext{b}{The $\chi ^{2}$ of this solution are high, also the 
$^{13}$CO observations are inconsistent with these values (Figure 2a; 
\S 6), so this solution should be considered uncertain.}
\label{tbl4}
\end{deluxetable}

\begin{deluxetable}{ccc}
\tablenum{5}
\tablewidth{0pt}
\tablecaption{Off-Center Line Ratios}
\tablehead{
\colhead{Galaxy} &\colhead{Location} &\colhead{CO(3-2)/CO(1-0)}}
\startdata
He 2-10 & (20,-10) & $0.35 \pm 0.1$ \\  
   (D)     & (20,10) & $0.30 \pm 0.1$ \\  
   (C)     & (10,-10) & $0.35 \pm 0.1$ \\  
NGC 5253& (10,-10) & $<0.7$ \\  
        & (20,20) & $<0.6$ \\  
        & (15,25) & $<1.2$ \\  
\enddata
\tablecomments{Ratios are based on the CO(1-0) interferometer maps at
22$^{''}$ resolution, and have been corrected for resolved out flux (He
2-10: 82\%; Kobulnicky et al. 1995; NGC 5253: 50\%; Turner et
al. 1997).  Uncertainties for the CO(3-2) data are given in Table 3,
while the interferometric data is assumed to have an uncertainty of
20\%.  In the case of NGC 5253 the ratio is of peak T$_{mb}$ since
these three CO(3-2) pointings are non-detections: quoted values are
3$\sigma$ upper limits.}
\label{tbl5}
\end{deluxetable}

\begin{deluxetable}{lcccccccc}
\tablenum{6}
\tablewidth{0pt}
\tablecaption{Gas Masses}
\tablehead{
\colhead{Galaxy} & \colhead{T$_{ex}$} &\colhead{$\frac{X}{X_{Gal}}$
\tablenotemark{a}}
& \colhead{$\frac{M_{vir}}{sin^{2}i}$\tablenotemark{b}}
&\colhead{$\frac{M_{thin}}{[CO/H_{2}]}$}
&\colhead{M$_{mol}$}&\colhead{$M_{mol} \over M_{vir}$
$sin^{2} i$ } & \colhead{$M_{mol}\over M_{D}$} &\colhead{$M_{gas} \over 
M_{vir}$$sin^{2} i$\tablenotemark{c}} \\
\colhead{} & \colhead{(K)} & \colhead{} &
\colhead{(10$^{6}$ M$_{\odot}$)}& \colhead{(10$^{6}$ M$_{\odot}$)}&
\colhead{(10$^{6}$ M$_{\odot}$)}&\colhead{}& \colhead{}&
\colhead{}}
\startdata
He 2-10&11&1.0&330&1.0&140&0.42&475 &1.4\\
NGC 5253&15&7.0&94&0.004&37&0.39& 680&2.5 \\
NGC 1569&50&5.7&26 &0.007&4.6&0.18& 140&3.1 \\
NGC 3077&50&0.9&93&0.04&4.4&0.05& 45 &8.6 \\
Haro 2&6&3.5&1100&2.1&750&0.68& 2300 &1.1 \\
Haro 3&50&3.5&430&0.25&110&0.26& 790 &1.7 \\
II Zw 40&\nodata&6.1&\nodata &\nodata&$<0.4$&\nodata&$<$830&\nodata\\
Mrk 86&\nodata &7.7&\nodata &\nodata&$<0.3$&\nodata& $<$340&\nodata\\
\enddata
\tablenotetext{a}{Using the single-dish $X$ vs. metallicity relation of 
Arimoto et al. 1996: $log\left (X \over X_{Gal} \right ) = 
-[O/H]~+~8.93$}
\tablenotetext{b}{Using: M$_{vir}$ = 210 $R_{pc}~\Delta V_{1/2}^{2}
~M_{\odot}$, (MacLaren, Richardson \& Wolfendale 1988)}
\tablenotetext{c}{M$_{gas}$ = M$_{mol}$ + M$_{HI}$, (see Table 1)}
\label{tbl6}
\end{deluxetable}


\begin{thebibliography}{}
\bibitem[Aalto et al. (1995)]{ABBJ95}Aalto, S., Booth, R. S., Black, J. H. 
\& Johansson, L. E. B. 1995, \aap, 300, 369
\bibitem[Arimoto, Sofue and Tsujimoto (1996)]{AST96}Arimoto, N., Sofue, Y. 
\& Tsujimoto, T. 1996, \pasj, 48, 275
\bibitem[Arnault et al. (1988)]{ACCK88}Arnault, P., Casoli, F., Combes, 
F., \& Kunth, D. 1988, \aap, 205, 41
\bibitem[Arp and Sandage (1985)]{AS85}Arp, H. \& Sandage, A. 1985, \aj, 90, 
1163
\bibitem[Baas, Israel and Koornneef (1994)]{BIK94}Baas, F., Israel, F. P. 
\& Koornneef, J. 1994, \aap, 284, 403
\bibitem[Beck and Kovo (1998)]{BK98}Beck, S. C. \& Kovo, O. 1998, \aj,
 117, 190
\bibitem[Becker, Schilke and Henkel (1989)]{BSH89}Becker, R., Schilke, P. 
\& Henkel, C. 1989, \aap, L211, 19
\bibitem[Braine and Combes (1992)]{BC92}Braine, J. \& Combes, F. 1992, 
\aap, 264, 433 
\bibitem[Conti (1991)]{C91}Conti, P. S. 1991, \apj, 377, 115
\bibitem[Conti and Vacca (1994)]{CV94}Conti, P. S. \& Vacca, W. D. 1994, 
\apj, 473, L97
\bibitem[Cottrell (1976)]{C76}Cottrell, G. A. 1976, \mnras, 174, 455
\bibitem[De Jong, Chu and Dalgarno (1975)]{DCD75}De Jong, T., Chu, S.-I. 
\& Dalgarno, A. 1975, \apj, 199, 69
\bibitem[Devereux et al. (1994)]{DYSNY94}Devereux, N., Yoshiaki, T., 
Sanders, D. B., Nakai, N. \& Young, J. S. 1994, \aj, 107, 2006
\bibitem[Eckart et al. (1990)]{E90} Eckart, A., Downes, D., Genzel, 
R., Harris, A. I., Jaffe, D. T.,\& Wild, W. 1990, ApJ, 348, 434
\bibitem[Elmegreen (1989)]{E89}Elmegreen, B. G. 1989, \apj, 338, 178
\bibitem[Frerking, Langer and Wilson (1982)]{FLW82}Frerking, M. A., 
Langer, W. D. \& Wilson, R. W. 1982, \apj, 262, 59
\bibitem[Gallagher and Hunter (1984)]{GH84}Gallagher, J. S. \& Hunter, 
D. A. 1984, \araa, 22, 37
\bibitem[Garcia-Burillo, Guelin and Cernicharo (1993)]{GGC93}Garcia-Burillo, 
S., Guelin, M. \& Cernicharo, J. 1993, \aap, 274, 123
\bibitem[Garnett (1990)]{G90}Garnett, D. R. 1990, \apj, 363, 142
\bibitem[Goldreich and Kwan (1974)]{GK74}Goldreich, P. \& Kwan, J. 
1974, \apj, 189, 441
\bibitem[Gonzalez Delgado et al. (1997)]{GLHC97}Gonzalez Delgado, R. M., 
Leitherer, C., Heckman, T. \& Cervino, M. 1997, \apj, 483, 705
\bibitem[Greve et al. (1996)]{GBJM96}Greve, A., Becker, R., Johansson, 
L. E. B. \& McKeith, C. D. 1996, \aap, 312, 391
\bibitem[Heckman (1980)]{H80}Heckman, T. M. 1980, \aap, 87, 142
\bibitem[Hodge (1971)]{H71}Hodge, P. W. 1971, \araa, 9, 35 
\bibitem[Hurt et al. (1993)]{HTHM93}Hurt, R. L., Turner, J. L., Ho, 
P. T. P. \& Martin, R. N. 1993, \apj, 404, 602
\bibitem[Israel (1986)]{I86}Israel, F. P. 1986, \aap, 168, 369
\bibitem[Israel (1988)]{I88}Israel, F. P. 1988, \aap, 194, 24
\bibitem[Israel and de Bruyn (1988)]{ID88}Israel, F. P., de Bruyn, 
A. G. 1988, \aap, 198, 109
\bibitem[Israel et al. (1986)]{IDVD86}Israel, F. P., De Gwaauw, 
Th., Van de Stadt, H. \& De Vries, C. P. 1986, \apj, 303, 186
\bibitem[Israel and van Driel (1990)]{IVD90}Israel, F. P., \& van Driel, 
W. 1990, \aap, 236, 323
\bibitem[Klein, Weiland and Brinks (1991)]{KWB91}Klein, U., Weiland, H., 
\& Brinks, E. 1991, \aap, 246, 323
\bibitem[Kobulnicky et al. (1995)]{KDSHC95}Kobulnicky, H. A., Dickey, 
J. M., Sargent, A. I., Hogg, D. E. \& Conti, P. S. 1995, \aj, 110, 116
\bibitem[Kobuknicky and Johnson (1999)]{KJ99}Kobulnicky, \& H. A., 
Johnson, K. 1999, \apj, 527, 154
\bibitem[Kobuknicky, Kennicutt and Pizagno (1998)]{KKP98}Kobulnicky, H. A., 
Kennicutt, R. C., \& Pizagno, J. L. 1998, \apj, 514, 544
\bibitem[Kobulnicky and Skillman (1995)]{KS95}Kobulnicky, H. A., \& Skillman, 
E. D. 1995, \apjl, L454, 121
\bibitem[Kobulnicky and Skillman (1997)]{KS97}Kobulnicky, H. A., \& Skillman, 
E. D. 1997, \apj, 489, 636
\bibitem[Kunth and Jobert (1985)]{KJ85}Kunth, D., \& Jobert, M. 1985, 
\aap, 142, 411
\bibitem[Lequeux et al. (1994)]{LLPRBR94}Lequeux, J., Le Bourlot, J., 
Pineau Des Forets, G., Roueff, E., Boulanger, F. \& Rubio, M. 1994, \aap, 
292, 371
\bibitem[MacLaren, Richardson and Wolfendale (1988)]{MRW88}MacLaren, I., 
Richardson, K. M. \& Wolfendale, A. W. 1988, \apj, 333, 821
\bibitem[Maloney and Black (1988)]{MB88}Maloney, P. \& Black, J. H. 1988, 
\apj, 325, 389
\bibitem[Maloney and Wolfire (1997)]{MW97}Maloney, P., \& Wolfire, 
M. G. 1997, in IAU Symp 170: CO: Twenty-Five Years of Millimeter-Wave 
Spectroscopy, ed. W. B. Latter et al. (Dordrecht:Kluwer), 299 
\bibitem[Marconi, Matteucci and Tosi (1994)]{MMT94}Marconi, G., 
Matteucci, F., \& Tosi, M. 1994, \mnras, 270, 35
\bibitem[Mauersberger et al.(1999)]{MHWS99}Mauersberger, R., Henkel, C., 
Walsh, W. \& Schulz, A. 1999, \aap, 341, 256
\bibitem[Meier and Turner (1998)]{MT98}Meier, D. S., \& Turner, 
J. L. 1998,in: K.A. van der Hucht, G. Koenigsberger \& P.R.J.  
Eenens (eds.), Wolf-Rayet Phenomena in Massive Stars and Starburst 
Galaxies, Proc. IAU Symp. No. 193 (San Francisco: ASP), 746
\bibitem[Meier and Turner (2000)]{MT00}Meier, D. S. \& Turner, 
J. L. 2000, in prep.
\bibitem[Meier, Turner and Hurt (2000)]{MTH00}Meier, D. S., Turner, J. L., 
\& Hurt, R. L. 2000, \apj, 531, 200.
\bibitem[Melisse and Israel (1994)]{MI94}Melisse, J. P. M. \& Israel, 
F. P. 1994, A\&AS, 103, 391
\bibitem[Meurer et al. (1995)]{MHLKRG95}Meurer, G. R., Heckman, T. M., 
Leitherer, C., Kinney, A., Robert, C. \& Garnett, D. R. 1995, \aj, 110, 
2665
\bibitem[Mochizuki et al. (1994)]{MET94}Mochizuki, K., et al. 1994, \apjl, 
430, L37
\bibitem[Niklas et al. (1995)]{NKBW95}Niklas, S., Klein, U., Braine, 
J., \& Wielebinski, R. 1995, A\&AS, 114, 21
\bibitem[Pak et al. (1998)]{PJVJB98}Pak, S., Jaffee, D. T., van Dishoeck, 
E. F., Johansson, L. E. B., \& Booth, R. S. 1998, \apj, 498, 735
\bibitem[Petitpas and Wilson (1998)]{PW98}Petitpas, G. R. \& Wilson, 
C. D. 1998, \apj, 496, 226
\bibitem[Reakes (1980)]{R80}Reakes, M. 1980, \mnras, 192, 297
\bibitem[Rohlfs and Wilson (1996)]{RW96}Rohlfs, K. \& Wilson, T. L. 1996, 
Tools of Radio Astronomy, 2nd Ed., (Springer-Verlag:Berlin) 
\bibitem[Rubio, Lequeux and Boulanger (1993)]{RLB93}Rubio, M., Lequeux, J. 
\& Boulanger, F. 1993, \aap, 271, 9
\bibitem[Sage et al. (1992)]{SSLH92}Sage, L. J., Salzer, J. J., 
Loose, H.-H. \& Henkel, C. 1992, \aap, 265, 19
\bibitem[Sandage (1994)]{S94}Sandage, A. 1994, \apjl, L423, 13
\bibitem[Sanders et al. (1993)]{STSWZ93}Sanders, D. B., Tilanus, 
R. P. J., Scoville, N. Z., Wang, Z., \& Zhou, S. 1993, in Back to 
the Galaxy, ed. F. Verter (Kluwer, Dordrecht), 21
\bibitem[Schaerer, Contini and Pindao (1999)]{SCP99}Schaerer, D., Contini, T. 
\& Pindao, M 1999, A\&AS, 136, 35
\bibitem[Skillman, Kennicutt and Hodge (1989)]{SKH89}Skillman, E. D., 
Kennicutt, R. C. \& Hodge, P. W. 1989, \apj, 347, 875
\bibitem[Steel et al. (1996)]{SSMRM96}Steel, S. J., Smith, N., Metcalfe, 
L., Rabbette, M. \& McBreen, B. 1996, \aap, 311, 721
\bibitem[Stil and Israel (1998)]{SI98}Stil, J. M. \& Israel, F. P. 1998, 
\aap, 337, 64
\bibitem[Strong et al. (1988)]{SET88}Strong et al. 1988, \aap, 207, 1
\bibitem[Tacconi and Young (1987)]{TY87}Tacconi, L. J. \& Young, J. S. 
1987, \apj, 322, 681
\bibitem[Tammann and Sandage (1968)]{TS68}Tammann, G., A., \& Sandage, 
A. 1968, \apj, 151, 825
\bibitem[Taylor et al. (1999)]{THKG99}Taylor, C. L., H\"{u}ttemeister, 
S., Klein, U., \& Greve, A. 1999, \aap, 349, 424
\bibitem[Taylor, Kobulnicky and Skillman (1998)]{TKS98}Taylor, C. L., 
Kobulnicky, H. A. \& Skillman, E. D. 1998, \aj, 116, 2746
\bibitem[Thronson and Telesco (1986)]{TT86}Thronson, H. A. \& Telesco, 
C. M. 1986, \apj, 311, 98
\bibitem[Thuan and Martin (1981)]{TM81}Thuan, T. X. \& Martin, G. E. 
1981, \apj, 247, 823
\bibitem[Turner, Beck and Hurt (1997)]{TBH97}Turner, J. L., Beck, S. C. \& 
Hurt R. L. 1997, \apjl, L474, 11
\bibitem[Turner, Ho and Beck (1998)]{THB98}Turner, J. L., Ho, P. T. P., \& 
Beck, S. C. 1998, \aj, 116, 1212
\bibitem[Turner, Hurt and Hudson (1993)]{THH93}Turner, J. L., Hurt R. L., \& 
Hudson, D. Y. 1993, \apjl, L413, 19
\bibitem[Vacca and Conti (1992)]{VC92}Vacca, W. D. \& Conti, P. S. 1992, \apj,
 401, 543
\bibitem[van Dishoeck and Black (1988)]{VB88}van Dishoeck, E. F., \& 
Black, J. H. 1988, \apj, 334, 771
\bibitem[Verter and Hodge (1995)]{VH95}Verter, F. \& Hodge, P. 1995, \apj, 
446, 616
\bibitem[Waller (1991)]{W91}Waller, W. H. 1991, \apj, 370, 144
\bibitem[Warin, Benayoun and Viala (1996)]{WBV96}Warin, S., Benayoun, J. J., 
\& Viala, Y. P. 1996, \aap, 308, 535 
\bibitem[Wiklind and Henkel (1989)]{WH89}Wiklind, T. \& Henkel, C. 1989, 
\aap, 225, 1
\bibitem[Wild et al. (1992)]{WHEGGJRS92}Wild, W., Harris, A. I., Eckart, 
A., Genzel, R., Graf, U. U., Jackson, J. M., Russell, A. P. G. \& 
Stutzki, J. 1992, \aap, 265, 447
\bibitem[Wilson (1995)]{W95}Wilson, C. D. 1995, \apjl, L448, 97
\bibitem[Wilson, Howe and Balogh (1999)]{WHB99}Wilson, C. D., Howe, J. E, \& 
Balogh, M. L. 1999, \apj, 517, 174
\bibitem[Wilson et al. (1988)]{WSFMS88}Wilson, C. D., Scoville, N. Z., 
Freedman, W. L., Madore, B. F., \& Sanders, D. B. 1988, \apj, 333, 611
\bibitem[Wilson, Walker and Thornley (1997)]{WWT97}Wilson, C. D., Walker, 
C. E. \& Thornley, M. D. 1997, \apj, 483, 210
\bibitem[Wolfire et al. (1990)]{WTH90} Wolfire, M. G., Tielens, 
A. G. G. M. \& Hollenbach, D. J. 1990, \apj, 358, 116 
\bibitem[Young, Gallagher and Hunter (1984)]{YGH84}Young, J. S., Gallagher, 
J. S. \& Hunter, D. A. 1984, \apj, 276, 476
\bibitem[Young and Scoville (1991)]{YS91}Young, J. S. \& Scoville, N. 
Z. 1991, \araa, 29, 581

\end{thebibliography}
\end{document}